%% file: autoreferat.tex
\documentclass[twoside,11pt]{article}

\include{packages}
\include{macros}
\include{ludmacros}

\begin{document}

\include{title}

\input{00}
\input{01}
\input{02}

\input{03}
\input{04}

\input{05}

\addcontentsline{toc}{section}{References}

\bibliographystyle{apalike}
\bibliography{bibliography}



\end{document}

%% file: packages.tex
\usepackage[utf8]{inputenc}

\usepackage{amssymb}        %
\usepackage{amsfonts}       
\usepackage{amsmath}
\usepackage{amsthm}         
\usepackage{mathrsfs}
\usepackage{bbding}         
\usepackage{bm}             
\usepackage{graphicx}       
\usepackage{fancyvrb}       
\usepackage{natbib}         
\usepackage{dcolumn}        
\usepackage{booktabs}       
\usepackage{paralist}       
\usepackage[usenames]{xcolor}  
\usepackage[british]{babel}
\usepackage{tcolorbox}

\usepackage[a-2u]{pdfx}

\usepackage{lmodern}

\usepackage{floatrow}
\usepackage{enumerate}
\usepackage{changepage}
\usepackage{caption}
\usepackage{geometry}
\usepackage{a4wide}
\usepackage{fancyhdr}
\usepackage{kantlipsum}
\usepackage[all]{nowidow}
\usepackage[labelfont=bf]{caption}
\usepackage[rightcaption]{sidecap}
\usepackage{microtype}
\usepackage{pdfpages}

\sidecaptionvpos{figure}{c}
\graphicspath{{img/}}
\def\Department{Astronomical Institute of Charles University}

\def\StudyProgramme{Theoretical Physics, Astronomy and Astrophysics}
\def\StudyBranch{P4F1}

%% file: macros.tex
\makeatletter
\def\@makechapterhead#1{
  {\parindent \z@ \raggedright \normalfont
   \Huge\bfseries \thechapter. #1
   \par\nobreak
   \vskip 20\p@
}}
\def\@makeschapterhead#1{
  {\parindent \z@ \raggedright \normalfont
   \Huge\bfseries #1
   \par\nobreak
   \vskip 20\p@
}}
\makeatother

\theoremstyle{plain}

\theoremstyle{plain}

\theoremstyle{remark}

\DefineVerbatimEnvironment{code}{Verbatim}{fontsize=\small, frame=single}



%% file: title.tex
\pagestyle{empty}

\begin{center}
\leavevmode\vskip1cm

\begin{figure}
\includegraphics[width=\textwidth]{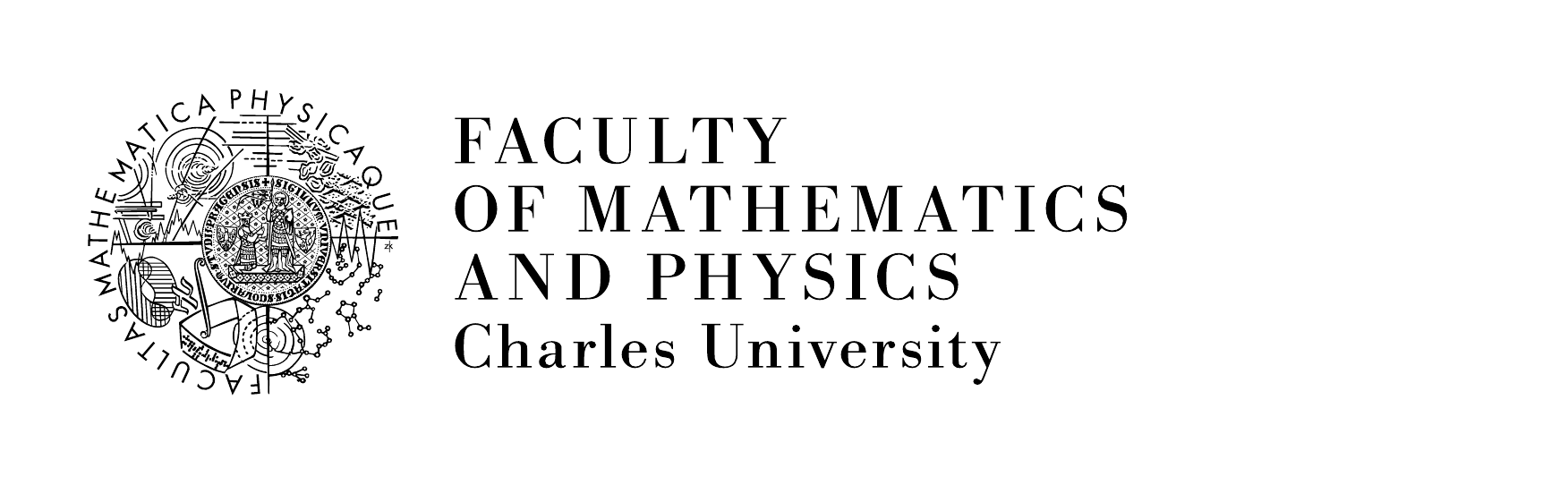}
\end{figure}

\vspace{-0.8cm}

Abstract of the doctoral thesis

\vspace{1.8cm}

{\Large\bf Chemical evolution of galaxies with an environment-dependent stellar initial mass function}

\vspace{1.8cm}

{\Large\bf Yan Zhiqiang}

\vspace{1.8cm}

\Department

\vspace{2cm}

{
\centerline{\vbox{\halign{\hbox to 0.45\hsize{\hfil #}&\hskip 0.5em\parbox[t]{0.45\hsize}{\raggedright #}\cr
Supervisor of the doctoral thesis:&Prof. Dr. Pavel Kroupa \cr
\noalign{\vspace{2mm}}
Study programme:&\StudyProgramme \cr
\noalign{\vspace{2mm}}
Study branch:&\StudyBranch \cr
}}}}

\vspace{2cm}
\vfill

Prague 2021

\end{center}


\newpage

\pagestyle{plain}
\pagenumbering{roman}

\leavevmode

\noindent
The doctoral thesis was done at the \Department\ between 2017 and 2021.

\vskip1\baselineskip

\begin{tabular}{@{}p{6.8cm}l}
{\bf PhD student:}  & Yan Zhiqiang \\[20pt]

{\bf Department:} & \Department \\
                & V Hole\v{s}ovi\v{c}k\'{a}ch~2 \\
                & 180\,00 Praha~8 \\
                & Czech Republic \\[20pt]

{\bf Supervisor:} & Prof. Dr. Pavel Kroupa \\
                & Astronomical Institute of Charles University \\
                & V Hole\v{s}ovi\v{c}k\'{a}ch~2 \\
                & 180\,00 Praha~8 \\
                & Czech Republic \\[20pt]

{\bf Referees:} & Prof. Dr. Daniel Thomas \\
                & Institute of Cosmology and Gravitation \\
                & University of Portsmouth \\
                & Burnaby Road \\
                & Portsmouth, UK \\
                & PO1 3FX \\[12pt]
                & Dr.~Ignacio Mart\'in-Navarro \\
                & Instituto de Astrofisica de Canarias \\
                & Via Lactea S/N, 38200 \\
                & La Laguna, Tenerife \\
                & Spain \\
\end{tabular}

\noindent

\vskip\baselineskip

\noindent
The defence of the thesis will be held on Friday, December 17, 2021, at 10:00 (Central European Time) in front of the defence committee for the doctoral study branch \StudyBranch\ \StudyProgramme, via online Zoom meeting due to travel restrictions.\\ZOOM Meeting ID: 955 9318 4633\\Passcode: 774984

\vspace*{8pt}

\noindent
The thesis can be obtained in printed or electronic version through the Student Affairs Department of the Faculty of Mathematics and Physics of Charles University, Ke Karlovu~3, Praha~2.

\vspace*{8pt}

\begin{tabular}{@{}p{6.8cm}l}
{\bf Chairman of the programme \StudyBranch:}
                & prof. RNDr. Pavel Krtous, Ph.D. \\
                & Institute of Theoretical Physics MFF UK \\
                & V Hole\v{s}ovi\v{c}k\'{a}ch~2 \\
                & 180\,00 Praha~8 \\
                & Czech Republic \\
\end{tabular}


\newpage

\tableofcontents

\vfill

\section*{Acknowledgements}

Part of my doctoral study is supported by the China Scholarship Council (CSC, file number 201708080069). The development of our chemical evolution model applied in this work benefited from the International Space Science Institute (ISSI/ISSI-BJ) in Bern and Beijing, thanks to the funding of the team ``Chemical abundances in the ISM: the litmus test of stellar IMF variations in galaxies across cosmic time'' (Donatella Romano and Zhi-Yu Zhang).

\pagestyle{plain}
\pagenumbering{arabic}
\setcounter{figure}{0}
\setcounter{page}{1}

\newpage

%% file: 00.tex
\section*{Preface}
\addcontentsline{toc}{section}{Preface}

Modern astrophysics has gone through two major revolutions. One is the Monte Carlo method and Bayes' theorem. The other is the advances in computational power and numerical optimisation routines. As a result, scientific development has become dependent more on pure and parallel computation rather than discursive and linearly ordered reasoning. The conclusions are given more precise and quantitative but not necessarily more helpful when it is buried in implicit assumptions, difficult to compare with inconsistent peer results.

The subject of the present work, that is, the study of galaxy chemical evolution (GCE), has benefited from and, in the meantime, is a victim of modern astrophysics development. The numerical GCE study depends heavily on complicated galaxy models with many parameters describing physical processes not independently well constrained but only justified by the fact that the model as a whole can reproduce the observed feature of galaxies. The GCE study aims to find the most likely formation and evolution history of galaxies that reproduce the chemical abundances we observe today, and by doing so, achieves a better understanding of how galaxies form. However, this methodology is greatly weakened by our capability to accurately measure the chemical abundances of the real galaxies and the limitations of our galaxy model. 

On one hand, the measured galaxy abundance that a GCE code aims to reproduce is not certain at all. The galaxy chemical abundance given by a stellar population synthesis (SPS) code depends on the assumed spectrum of single stars and the distribution of the stellar masses, ages, and element abundances. There is a limited number of spectrum models with different stellar initial element abundances, evolving to different phases in the lifetime of a star, and affected by different degrees of dust attenuation. These spectrum models are either calibrated by local stars close to our Sun, which have similar abundances and do not cover every stellar evolution phase or are predicted by theoretical models that are difficult to validate. It is true that different stellar spectrum libraries provide spectra similar enough to not cause a significant difference in the results of the SPS but systematic errors and confirmation bias cannot be excluded. If the spectrum model is uncertain, then there is a risk when comparing metal abundance measurements based on different spectral features and even between stars and gas.

On the other hand, there are large uncertainties in a GCE model. When and how stars form and what elements do they eject during their final explosion are under debate. The simulation of the hydrodynamic evolution of the gas in a galaxy is not accurate. The physical processes such as gas mixing, cooling, accretion, and baryonic feedback in GCE models are still represented by parameters that are not well constrained, let along with the gas with different temperatures and abundances of different elements. In fact, it is not a surprise that we can find more degrees of freedom in a GCE model than the number of observational constraints because we observe only the present-day snapshot of a galaxy while the GCE model simulates its evolution. The more complex a model becomes and the more free parameters (or not well-constrained parameters) it includes to explain a given set of observations, the less predictive power it has.

In particular, the standard GCE model has originally been developed to explain and, therefore, being calibrated by the observation of stars in the Milky Way galaxy (MW). The complex formation history of our disc galaxy results in multiple stellar populations with significantly different abundance features that require an equally complex GCE model to explain. However, for the extra-galactic galaxies, the information we have accumulated is much less and many of them form in environments nothing like the MW. The consequence is that the complex GCE model developed over the years with parameters fine-tuned to fit the MW observations is not directly applicable to the extra-galactic galaxies. The value of the many free parameters has not been validated in a different environment nor do we have enough independent observational constraints to do so. The solution to the current situation is to apply a much more simplified GCE model, a basic version that balances the number of free parameters and observational constraints, and starts from the beginning to first explain the global properties of the extra-galactic galaxies. Basically, we need to climb the ladder again from a different starting point to reach a different roof. We need to first adjust the most influential zero-order parameters before we trying to fit the higher-order details.

For example, it has been a long-standing problem to understand the fast formation and early enrichment of the massive elliptical galaxies applying the standard GCE models. Their stellar population appears to be too metal-rich and $\alpha$-element\footnote{$\alpha$-elements, including O, Mg, Si, S, and Ca, are synthesised by core-collapse supernovae through $\alpha$-particle capture.} enhanced \citep{2005ApJ...621..673T,2019A&A...632A.110Y}. The X-ray observation of the gas surrounding the galaxy clusters supports this picture -- far more metal exists than the amount a standard GCE model can produce (\citealt{2014MNRAS.444.3581R,2017MNRAS.470.4583U,2021A&A...646A..92G}, review: \citealt[their section 2.4]{2021Univ....7..208G}). Complex galactic feedback models have been developed in order to solve this mystery but failed \citep{2017MNRAS.466L..88D,2017MNRAS.464.4866O} because people are trying to reach a different roof without climbing down their current ladder.
Problems exist also on the lower end of the galaxy mass distribution. The ultra-faint dwarf galaxies appear to have an abnormally low star formation efficiency \citep{2014MNRAS.441.2815V,2015MNRAS.446.4220R} and abundance ratios difficult to understand within the standard GCE model \citep{2018ApJ...852...50F,2020A&A...642A.176T,2021ApJ...910..114M}. It is shown in \citet{2009ApJ...706..516P} and one of our studies below (Section~\ref{Chapter dwarf}) that these abnormal features are only apparent when one applies the standard GCE model overfitted by MW observations. Then the only question is, what is the most important component of a GCE that we need to reconsider when studying the extra-galactic galaxies? 

Unarguably, the stellar initial mass function (IMF) in the GCE model needs to be examined with caution, that is, how many low-mass stars and how many massive stars form in a stellar population. It has been studied in detail through observation and computational models on how stars evolve differently due to the difference in their initial masses. Low-mass stars with a mass of less than about eight solar masses (denoted by $M_\odot$ throughout this thesis) end up as white dwarfs which then have a chance to ignite as a thermonuclear supernova if they obtain enough additional mass from a companion star. The best-known object in this class, the Type Ia supernova (SNIa, we use this term interchangeably with thermonuclear supernova throughout this thesis), produces more than half of the iron peak elements in the universe over a Gyr to 10 Gyr timescale after the formation of their progenitor stellar population. Massive stars above about 8 $M_\odot$, on the other hand, are likely to explode as a core-collapse supernovae (CCSN) within 40 Myr, a blink of an eye in galaxy evolution. CCSN distribute most of the $\alpha$ elements to the galaxy and may leave a neutron star or black hole stellar remnant. Therefore, the GCE involving the abundance of different elements in a galaxy depends strongly on the mass distribution of the stars, that is, the IMF. This is indeed the standing ground of our ladder for any GCE research.

The standard GCE model has been assuming that the IMF for all stellar populations is the same -- the universal and invariant IMF. This idea is supported by the observation of the local universe, where it is found that stellar populations in vastly different environments across several orders of magnitude of temperature and density have indistinguishable IMFs \citep{2002Sci...295...82K,2010ARA&A..48..339B}. This is partially due to the fact that the measurement of the IMF is difficult and has large uncertainties. With new IMF indicators and better telescopes for observing extra-galactic galaxies that live in more extreme environments, a \textit{systematic} variation of the IMF has been discovered (see the overwhelming evidence listed in Section~\ref{sec IMF}) and is consistently described by the integrated-galactic IMF (IGIMF) theory \citep[and references therein]{2017A&A...607A.126Y}. This recent development of the IGIMF theory needs to be convolved with the GCE studies.

The IMF variation in GCE is the theme of my doctoral study where a new GCE model with an environment-dependent IMF and a high temporal resolution is developed particularly for this purpose. The implications of this important model paradigm shift are discussed. For example, how the constraints are changed regarding the physics of stellar evolution and formation, the yield and frequency of core-collapse and Type Ia supernovae, the formation history of galaxies and the recycling of gas. We explore a vast parameter space of SFH, gas supplies, different IMF variation models and SNIa models to demonstrate their effects, utilizing the SIRRAH computer cluster of the Astronomical Institute of the Faculty of Mathematics and Physics of Charles University. The GCE results with the variable galaxy-wide IMF (gwIMF, calculated with the IGIMF theory) are compared side by side with the same model assuming an invariant canonical IMF.

In Section~\ref{sec IMF}, we introduce the development of the IMF theory, explore the literature on the IMF variation indicators and the pattern of how the IMF varies with the physical properties of the star-forming region. 
The mathematical formula of the IGIMF theory that account for the observed IMF variation is developed and programmed. Section~\ref{sec IGIMF} introduces the IGIMF theory and confronts the theoretical expectation with observation.
The IGIMF theory is implemented in the galaxy chemical evolution model in Section~\ref{sec GCE}. A publicly available python code, GalIMF, is developed to allow for the environment-depended IMF variation and custom applications.
Two realistic applications of the GalIMF are further explored. In Section~\ref{Chapter dwarf}, a case study to reproduce the observed property of a dwarf galaxy, Bo\"otes~I, demonstrates that the IGIMF theory explains the observation better than an invariant canonical IMF. The introduction of the new IMF formulation changes our understanding of the star formation efficiency of this ultra-faint dwarf galaxy (UFD). Similarly, when we apply the IGIMF theory to ETGs with different masses in Section~\ref{Chapter elliptical}, the estimation of their star formation timescales (SFTs) is different from the canonical model. The canonical model tends to result in extremely short SFT estimations for the massive ETGs, which are difficult to reconcile with hydrodynamical modelling of galaxy formation. This long-standing problem along with many observed ETG properties, such as their element abundance evolution, SNIa production efficiency, and M/L, from dwarf to the most massive ETGs, come together naturally and are straightforwardly explained, all are consequences of the application of the IGIMF theory to the monolithic collapse of post-Big Bang gas clouds. Finally, a summary of these studies and the conclusion is given.

This thesis is based on five refereed publications of which the author of this thesis is the first author (see attached list of publications). This series of publications are merged, restructured, and modified to give a linearly ordered reasoning.

%% file: 01.tex
\section{The variation of the stellar initial mass distribution}\label{sec IMF}

\subsection{Introduction: What is the IMF?}

\begin{figure}
    \center
    \includegraphics[width=9cm]{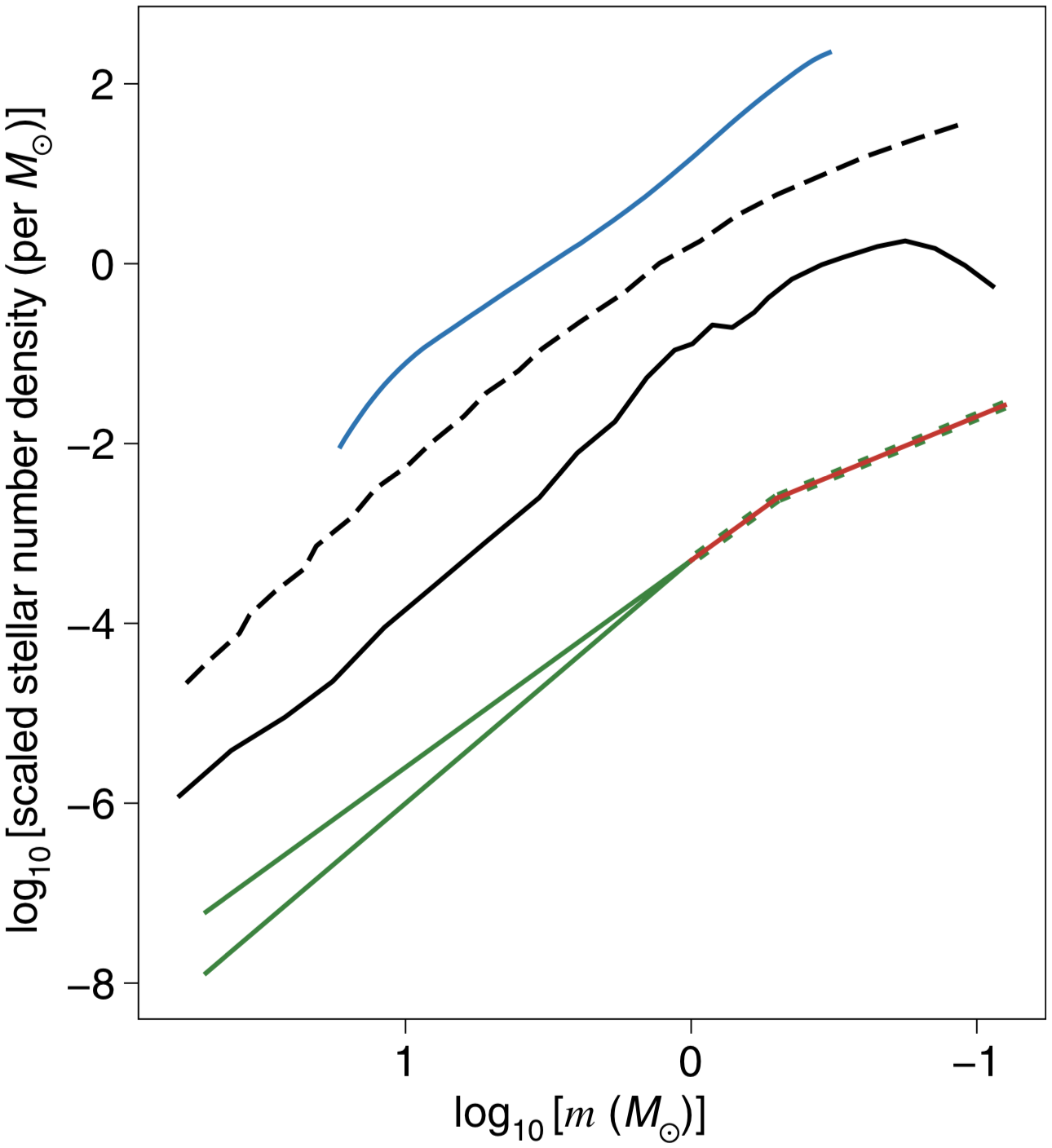}
    \caption{Figure taken from \citealt{2019NatAs...3..482K}. It shows the development history of IMF measurements. From top to bottom, the IMF determined by \citet{1955ApJ...121..161S} (blue), \citet{1979ApJS...41..513M} (dashed black), \citet{1986FCPh...11....1S} (solid black), and the present-day generally accepted Kroupa IMF (\citealt{1993MNRAS.262..545K} and \citealt{2001MNRAS.322..231K}, green and red, respectively, in near-exact agreement with the \citealt{2003PASP..115..763C} IMF). The IMFs are shifted vertically arbitrarily to demonstrate their differences.}
    \label{fig:field_star_IMF}
\end{figure}

The birth mass distribution of stars determines the observed properties of stellar populations and is critical to a wide range of astrophysical applications including star formation, stellar populations, galaxy evolution and more \citep{1998ASPC..142....1K}. This mass distribution of stars is described by their initial mass function (IMF), which is the function for the number of stars per stellar mass bin.
Since it is of crucial importance to achieve the correct IMF for the stellar population being studied, a huge amount of effort has been devoted to measuring the IMF since about a century ago with a growing number of review articles dedicated to it. See Fig.~\ref{fig:field_star_IMF} for a comparison of historical IMF measurements for the local field stars and \citet{2019NatAs...3..482K} for more details.

\subsection{Historical consensus of a universal invariant IMF}

Using resolved stellar populations in the local universe, different functional forms have been proposed to describe the IMF. Since 2000, a consensus has emerged for the massive end of the IMF that it is a power-law function. \citet{2001MNRAS.322..231K,2002Sci...295...82K} demonstrated that the observed IMF power-law index above a few solar masses shows a scatter that is smaller than random sampling would give (see \citealt{2013pss5.book..115K}, their Fig. 27) being consistent with the assumption that the IMF is universal and invariant. Therefore, it is generally accepted that the stellar IMF in star clusters is the universal canonical IMF, for example, the two-part power-law Kroupa IMF \citep{2001MNRAS.322..231K,2002Sci...295...82K,2010ARA&A..48..339B,2013pss5.book..115K}. The lack of scatter and the existence of the $m_{\rm max}$--$M_{\rm ecl}$ relation are interpreted by \citet{2013pss5.book..115K} to imply the IMF to be an optimal distribution function (and not a probability density) which is consistent with star formation being highly self-regulated.

\subsection{Theory}\label{sec IGIMF-IGIMF}

We still do not have a good idea on how the molecular clouds is able to overcome the thermal and magnetic pressure to condense itself over 20 orders of magnitude and transform violently yet inefficiency a few percent of its mass into stars \citep{2020MNRAS.493.2872C}. Traditional and influential ideas include hierarchical fragmentation, inside-out collapse, bimodal star formation, accretion from the protostellar disc and the coalescence of dense cores and previously formed stars within a crowded environment to form massive stars \citep{2004fost.book.....S}. 
Many teams are focusing on detailed hydrodynamical simulations studying the initial fragmentation of molecular clouds and trying to reach as high a resolution as possible in order to be able to capture the complex filamentary structure of molecular clouds. \citet{2012MNRAS.422.2246M}, applying N-body simulations, demonstrated that the IMF of globular clusters within the Milky Way is a function of their metallicity and mass.
Although much effort has been invested in constraining the shape and possible variation of the IMF, a comprehensive theory, describing star formation and resulting in an IMF being consistent with the observations both on star cluster and galaxy scales, has not been fully developed \citep{2019MNRAS.485.4852G}.

Nevertheless, the theoretical and computational explorations are useful in providing guidance on how different physical properties of the parental cloud might have an effect on the shape of the stellar IMF. 
A variable stellar IMF has been expected by theoretical studies for a long time since the star-forming environment should in principle affect the masses of stars that form, for example, through an ambient gas temperature, metallicity, density, and pressure dependences. The dense metal-poor cloud with an inefficient cooling process dominated by molecular hydrogen cooling should, in general, have a larger characteristic stellar mass. The early universe with a lower metallicity and a higher ambient temperature, should have yield higher Jeans masses and stellar masses. The extremely metal-poor Population III stars are predicted to have an IMF shifted to higher masses. Models considering pre-stellar condensations also suggest a top-heavy IMF (see below) in starburst clusters. 

Throughout this thesis, the non-canonical IMF shapes are described as top-light or top-heavy and bottom-light or bottom-heavy IMFs. The definition of these terms is clarified in \citet{2018A&A...620A..39J}. In short, the top and bottom refer to stars above and below $1~M_\odot$. The light and heavy indicate that there is a fewer or larger number of such stars than a similar population following the canonical \citet{2001MNRAS.322..231K} IMF.


\subsection{Observational evidence for the IMF variation}\label{sec IGIMF-Observation}

For star clusters, the primary difficulty in determining the IMF is that the measurement is affected strongly by the dynamical evolution of the star cluster. Low-mass stars are evaporated while massive stars are ejected and end their ``short'' life of only about 3 Myr when new stars can still form in the cluster and that most stars are still accreting masses and have not reached the main sequence phase yet. With careful account for the dynamical evolution and compare the observed radii of star clusters with models, \citet{2012MNRAS.422.2246M} argue that metal-poor clusters must have had top-heavy IMFs to reach their present-day sizes due to gas removal \citep{2007MNRAS.380.1589B}.
The ultra-compact dwarf galaxies, some considered as very massive star clusters, also show evidence of having a top-heavy IMF based on their dynamical V-band mass-to-light ratios and a large number of low mass X-ray binaries (LMXBs, \citealt{2009MNRAS.394.1529D,2010MNRAS.403.1054D} and \citealt{2012ApJ...747...72D}, although under debate).

The Milky Way field stars suggests a bottom-light galaxy-wide IMF (gwIMF) at early times when the Galactic metallicity is low.
Using numerical chemical evolution experiments, \citet{1990ApJ...365..539M} first point out that a slightly top-heavy IMF is required to explain the metallicity distribution of the Galactic bulge stars (confirmed by e.g. the [Mg/Ba] variation in the MW, where the two elements are produced mainly by different masses of stars \citealt{2021arXiv211101809H}).
Based on analysis of  carbon enhanced metal-poor stars (CEMP) in the Milky Way, \citet{2007ApJ...658..367K,2011MNRAS.412..843S,2013MNRAS.432L..46S} favour an IMF that changes from bottom-light to bottom-heavy. \citet{2009A&A...508.1359I} and \citet{2014ApJ...788..131L} suggest a similar scenario based on the binary fraction of CEMP stars.
The absence of low-mass stars (that should have a long life and survived till today) below $[Fe/H]=-4$ (e.g. \citealt{2020MNRAS.492.4986Y}) is also consistent with having a bottom-light IMF at low metallicities.

Among different galaxies, independent methods have shown that the gwIMF for massive stars appear to vary systematically, being different in star-forming regions with a more extreme environment. Specifically, \citet{1983ApJ...272...54K} and \citet{1994ApJ...435...22K} develop the classic method to estimate the effects of the variation of IMF, dust abundance, galactic SFH, and stellar evolution model on the colour and magnitude of the integrated galactic light. Using the Kennicutt method and with better data and more detailed modelling, \citet{2008ApJ...675..163H} demonstrate that, to their surprise, the colour-magnitude distribution for a statistical sample of SDSS galaxies cannot be explained by different SFHs but an IMF varies systematically with the luminosity of the galaxies. Indeed, \citet{2011MNRAS.415.1647G} study the different evolutionary paths a galaxy would take in the H$\alpha$ equivalent width and g--r colour plane with a sample of about 33000 galaxies from the Galaxy And Mass Assembly (GAMA) survey and find that the stellar IMF within galaxies has a strong variation with the galaxy star formation rate (SFR) by becoming increasingly top-heavy with increasing SFR.
This systematic variation extends to the low-SFR galaxies. \citet{1987A&A...185...33B} examine the the galactic H$\alpha$ and the far-ultraviolet light to study IMF variation. This method is revisited by \citet{2009ApJ...695..765M} and \citet{2009ApJ...706..599L} with a statistical sample for galaxies, suggesting that dwarf and low surface brightness galaxies are deficient in high-mass stars.
The same systematic variation extends to the starburst galaxies. Using C, O and N isotope abundance ratios
, \citet{2017MNRAS.470..401R} and \citet{2018Natur.558..260Z} manage to constrain the stellar IMF in galaxies where the intense star formation is still embedded in molecular clouds and cannot be observed at optical wavelengths.
The independent galaxy chemical evolution studies confirm the same picture. There are two major type of constraints we can get from a chemical evolution model, (i) the abundance-ratio--abundance relation and (ii) the abundance--time relation. The (i) relation is directly affected by the shape of the high-mass IMF and has been used to identify IMF variation (\citealt{2018ApJ...852...50F,2020A&A...637A..68Y,2021ApJ...910..114M,2021NatAs.tmp..200M}) while the (ii) relation is affected, in addition, by the star formation time scale, star formation efficiency, and the IMF for the low-mass stars (see \citealt{1990ApJ...365..539M,2020A&A...637A..68Y} and \citealt{2021A&A...655A..19Y}). We note that stellar population synthesis based on the present-day integrated light of a galaxy can potentially result into a population of stars with an unrealistic abundance--age distribution if the adopted IMF is incorrect. Therefore, it is encouraging to find a consistent description of the IMF variation from both the integrated light method and the chemical evolution method.

For the gwIMF of the low-mass stars, \citet{2003MNRAS.340.1317V} first identify a very dwarf-dominated IMF based on spectral synthesis of the galaxy. Later evidence has discovered a clear correlation between the mass of nearby early type galaxies (ETG) and their IMF slope, supporting a more bottom-heavy IMF for more massive galaxies (\citealt{2012ApJ...760...71C,2017ApJ...846..166L} among many). \citet{2015ApJ...803...87S} show that the non-universality of the low-mass end of the IMF is robust against the choice of the SSP model.
In recent years, pioneering studies using spatially resolved spectroscopy are now starting a new branch of research. The radial gradients of IMF sensitive features in a sample of 24 ETGs observed by the CALIFA survey in \citet{2015ApJ...806L..31M} find that the IMF is tightly related to metallicity, becoming more bottom-heavy toward metal-rich populations. The most recent study assessing the diversity of IMF and stellar population maps within the Fornax Cluster confirms the strong metallicity--IMF slope correlation \citep{2021A&A...654A..59M} with metal-rich galaxies forming more low-mass stars than expected by the canonical IMF.

The lensing studies, suffering from an uncertain amount and density profile of the dark matter component as well as differences in the subgrid feedback models \citep{2021MNRAS.tmp.2737M}, lead to different deductions concerning the bottom-heavy IMF than suggested by the above studies on dwarf-star-sensitive spectral features. Many studies \citep{2010ApJ...721L.163A,2013MNRAS.436..253B} prefer a bottom-heavy IMF, being consistent with spectral studies. On the other hand, \citet{2013MNRAS.434.1964S,2015MNRAS.449.3441S} and \citet{2017ApJ...845..157N} find that some nearby ETGs are only consistent with a canonical IMF, suggesting tension between dwarf-star indicators and lensing-mass constraints. Finally, \citet{2016MNRAS.459.3677L} reveal a large range of allowed IMF slopes, which may explain the above disagreement, but they need to invoke a significant dark matter component which may be stellar remnants from a top-heavy IMF (cf. \citealt[their fig. 6]{2019A&A...629A..93Y} and \citealt[their fig. 10]{2021A&A...655A..19Y}) rather than a dark matter component \citep{2015CaJPh..93..169K}.


Looking at higher redshifts, galaxies show the same character. \citet{2015ApJ...798L...4M} suggest that the IMF of a sample of 49 massive quiescent galaxies at $0.9 < z < 1.5$ is bottom-heavier for more massive galaxies and has remained unchanged in the last $\approx 8$ Gyr. \citet{2017A&A...600L...3C} provide the first direct constraints on the IMF at $z = 1.5$ in a lens, deducing a bottom-heavy IMF. At redshift about 4 to 7, \citet{2011ApJ...735L..34G} find a steep low-mass IMF slope using FAST spectral energy distribution (SED) fitting with a constant SFR and a Salpeter-type (single slope) IMF. \citet{2014MNRAS.444.2960D}, using a similar method on the galaxies in the same redshift range, find a significantly steeper low-mass IMF slope at higher redshift.

However, the deduced bottom-heavy IMF or even the canonical invariant IMF, if we assume the same IMF power-law index can extrapolate from the low-mass to high-mass part of the IMF and from the observed time to the past of the galaxy, cannot account for the Solar or even super-Solar metallicities of massive galaxies at high redshift as is explained by \citet{1998MNRAS.301..569L} and \citet{2013MNRAS.435.2274W}. The metal enrichment of the massive galaxies is too early and too fast \citep{2017MNRAS.470.4583U}. 
Indeed, galaxy clusters appear to form more metal elements than expected assuming a canonical IMF, suggesting a top-heavy IMF. More massive galaxy clusters have smaller stellar-mass fraction but a similar intracluster medium (ICM) metallicity, thus generating a major tension with the nucleosynthesis expectation and inflating the ratio of metal mass in ICM and in stars to extremely high values (up to $\approx$6, \citealt{2014MNRAS.444.3581R}). A canonical population of stars cannot produce this amount of metals that pollute the ICM \citep[their fig. 7]{2021Univ....7..208G}.
In line with these, semi-analytic cosmological models of galaxy formation demonstrate that a top-heavy IMF in starbursts is needed to adequately account for the number of sub-millimetre galaxies (SMG) at high redshifts while previous models with an invariant IMF can not \citep{2011RMxAC..40...21M,2016MNRAS.462.3854L}. Likewise, dust produced only by low-to-intermediate-mass stars falls a factor 240 short of the observed dust masses of SMGs in the semi-analytic model, supporting the need for higher supernova yields, substantial grain growth in the interstellar medium or a top-heavy IMF if dust is produced by both low-mass stars and supernovae and is not efficiently destroyed by supernova shocks \citep{2014MNRAS.441.1040R}.
X-ray observations sensitive to the high-mass IMF also show disagreements with the low-mass IMF trend. \citet{2017ApJ...835..183C} count the low-mass X-ray binaries per unit near-IR luminosity in seven nearby low-mass ellipticals and conclude that the slope for low-mass stars in the suggested bottom-heavy IMF must not extrapolate to high-mass stars if the IMF is time-invariant. Similarly, \citet{2017ApJ...841...28P} studied the low-mass X-ray binary populations of nine local ETGs and found that IMFs which become increasingly top-light with velocity dispersion are rejected. Thus the IMF seems to have to be both bottom- and top-heavy. 

These evidence requires that an excess number of massive stars were formed for the massive galaxies appear, at the present day, to have a steep IMF slope for their long-lived low-mass stars. 
If the IMF is more bottom-heavy for higher mass galaxies at low redshift and more top-heavy for higher SFR galaxies at high redshift, then the only possible scenario is that the IMF was top-heavy for starburst galaxies in the early universe and bottom-heavy at lower redshifts where the metallicity is high. As pointed out by \citet{2013MNRAS.435.2274W} in their toy model, such a scenario is in good agreement with various observational constraints on massive elliptical galaxies, such as age, metallicity, $\alpha$-enhancement, mass-to-light ratio or the mass fraction of the stellar component in low-mass stars.
\citet{2013MNRAS.435.2274W} and \citet{2015MNRAS.448L..82F} show that no time-independent IMF is capable to reproduce the full set of constraints on the stellar populations of massive ETGs. 

~~

\begin{tcolorbox}
Observation suggests that the IMF power-law indices for the low-mass and massive stars should have evolved independently as a function of different physical parameters. The IMF measurements for the low-mass stars do not give information on how the IMF varies for the massive stars.
\end{tcolorbox}

~~

\subsection{Summary}

There is always an infinite number of theories explaining any single observation but only one explaining most of the observations. The systematic variation of the galaxy-wide IMF is the simplest explanation for the vast variety of observations mentioned above.
But what is more important is the physical reason for the IMF variation and we want to know if the galaxy-wide IMF variation and the IMF variation on the star cluster scale can be explained self-consistently in a unified theory.

One likely solution is to associate the high-mass IMF with the galaxy-wide SFR, and the low-mass IMF with the galaxy-wide metallicity. 
From the downsizing relation (i.e. star formation duration inversely correlates with the mass of the galaxy, \citealt{2005ApJ...621..673T,2010MNRAS.404.1775T}) we know that the early-time SFR of a galaxy positively correlates with its mass. Therefore, massive galaxies formed in the early universe, with a high SFR and low metallicity, results in a bottom-light and top-heavy IMF. Such an IMF leads to an early and fast metal-enrichment process and, therefore, a super-solar average stellar metallicity for the galaxy \citep{2019A&A...629A..93Y}. The high metallicity then leads to a bottom-heavy population of stars formed at a later time when the earlier formed massive stars have already died out \citep[their fig. 9]{2021A&A...655A..19Y}. Thus the IMF shape of a galaxy varies at different times. 

A self-consistent formulation between the gwIMF and the IMF on star cluster scales can be achieved by the IGIMF theory (see below) where the galaxy-wide stellar population is the combination of all star clusters, each of these having a different IMF in dependence of their metallicity and mass.
In Section~\ref{sec IGIMF}, we introduce the IGIMF theory, its mathematical expression, and its observational tests.

%% file: 02.tex
\section{The IGIMF theory}\label{sec IGIMF}

\subsection{Overview}\label{sec IGIMF-IGIMF-overview}

The evident IMF variation on small scales (i.e. individual star-forming regions) and in galaxies needs to be consistently explained by a unified IMF theory. A promising proposal, the integrated galactic initial mass function (IGIMF) theory, was developed over the years for this purpose. Originally formulated in \citet{2003ApJ...598.1076K}, in which it was called the ``field-star IMF''\footnote{There is a difference between the best IMF estimation of the Galactic field-stars near the sun, with a high-mass power-law index of about $-2.7$ \citep{1995MNRAS.277.1491K}, and the best IMF estimation of the Galactic star clusters, where the field stars suppose to originate from, with a high-mass power-law index of about $-2.3$ \citep{2001MNRAS.322..231K}.}, the IGIMF theory predicted and explains the gwIMF variation. The fundamental insight underlying the IGIMF theory is that the empirical systematic variation of the gwIMF, which appears to correlate with galactic SFR and metallicity, has its origin from the variation of the IMF on a molecular-cloud core and embedded star cluster scale where a correlated star formation event\footnote{That is, a gravitationally driven collective process of transformation of the interstellar gaseous matter into stars in molecular-cloud overdensities on a spatial scale of about one pc and within about one Myr \citealt{2003ARA&A..41...57L,2013pss5.book..115K,2016AJ....151....5M}} happens. In other words, there exists a universal law of the star-cluster-scale IMF shape which leads to the various IMF shapes of different composite systems.

The IGIMF theory was originally based on a universal IMF on the star-cluster scale (i.e. in \citealt{2003ApJ...598.1076K}) and embedded cluster mass function (ECMF)\footnote{The mass of an embedded stellar cluster is the total initial mass of all the stars formed in that star cluster. The word embedded refers to the fact that the cluster is embedded in dense molecular cloud when they form. Then the mass of a star cluster decreases due to dynamical evolution.} with later adoption of a systematically varying IMF given by \citet{2012MNRAS.422.2246M} and a varying ECMF formulation suggested by \citet{2013MNRAS.436.3309W}, leading to a good description in explaining observations. There are three central axioms of the IGIMF theory which are formulated based on observational constraints (cf. \citealt{2017A&A...607A.126Y}). These axioms are:
\begin{itemize}
  \item Stars do not form in isolation but in groups (embedded star clusters with a mass as low as about $5~M_\odot$ as is observed by \citealt{2012ApJ...745..131K}) where the mass distribution of the groups follows the ECMF. It is known, observationally, that massive stars are created together with a dominant population of solar-type objects. The ejection of massive stars from the centre of a star cluster has been studied thoroughly. Dynamical simulations show that massive stars can move more than 50~pc away from the centre within 1~Myr and can even become blue stragglers at this larger radii \citep{2015ApJ...805...92O,2016A&A...590A.107O}. Observationally, nearly all observed isolated O stars can be traced back to their birth star cluster either along its line of velocity \citep{2012MNRAS.424.3037G} or with a two-step ejection mechanism described by \citet{2010MNRAS.404.1564P}. Recently using the Hubble Space Telescope, \citet{2017ApJ...834...94S} have specifically looked at seven massive stars thought previously, from Spitzer observations, to be isolated massive stars in the Large Magellanic Cloud. All of these seven stars turn out to be within substantial but compact clusters. The low-mass stars may have formed in low-mass stellar groups that got disrupted and dissolved immediately after their formation. Thus, it is impossible to prove or disprove whether they formed in groups. However, formation in groups is reasonable and natural for these low-mass objects because stars form in molecular cloud overdensities that contain much more mass than the mass of the least massive star. Observations indeed suggest that most and perhaps all the observed stars were formed in embedded clusters \citep{2003ARA&A..41...57L,2005ESASP.576..629K,2016AJ....151....5M}. With the ansatz that all stars form in embedded star clusters, all the stars in a galaxy are accounted for by adding all the embedded star clusters formed in that galaxy. This leads to the so-called ``IGIMF equation'' (Eq.~\ref{eq:IGIMF} below). We note that the calculation of the gwIMF of low-mass stars depends not only on the assumption that they all form in a star cluster but also on the adopted shape of the ECMF (our second axiom below) which is also not well constrained by the observation. Fortunately, the low-mass stars do not actively participate in the chemical and energy evolution of the galaxy compared to the massive stars. The variations at the low-mass end of the IMF have a much weaker effect on galaxy properties than variations at the high-mass end \citep{2014MNRAS.442.3138F}.  
  \item The star formation process is highly self-regulated. There is a deterministic relation between the mass of each star, the mass of the star cluster that contains these stars, and the mass of all star clusters formed in a correlated (10~Myr) star formation epoch (see the optimal sampling theory in \citealt{2013pss5.book..115K,2015A&A...582A..93S,2017A&A...607A.126Y}). It is supported by observation that the masses of stars are not randomly sampled from the IMF for a star cluster. For example, The north and south part of the Orion A cloud have roughly the same total amount of stars being formed but the massive stars only exist in the north part of the cloud where massive star clusters exist \citep{2012ApJ...752...59H,2013ApJ...764..114H,2016AJ....151....5M}. The cluster mass distribution is also not random \citep{2012ApJ...761..124G}.
  This axiom is important because if the formation of stars can be described by random sampling, then the integrated IMF given by the ``IGIMF equation'' for two population of star clusters would stochastically be the same as long as they have the same total mass. While the IGIMF theory assuming optimal sampling suggests that the integrated IMF of a population of stellar groups or clusters (e.g. the galaxy-wide IMF) depends on the mass distribution of the star clusters included in the integration \citep{2003ApJ...598.1076K}, which is consistent with the observation of stellar systems near (e.g. Orion) and far (e.g. dwarf galaxies, \citealt{2009ApJ...706..599L}).
  \item A correlated population of star clusters is formed in a short but non-negligible timescale. The typical timescale for the optimally-sampled galactic stellar population to form is about $\delta t = 10\,$Myr. The embedded star clusters formed within this timescale are correlated and optimally populate the ECMF of this 10~Myr formation epoch. Outside this timescale, the embedded star clusters are independent, that is, not populating the same ECMF. The time-scale $\delta t = 10\,$Myr, discussed in \citet{2004MNRAS.350.1503W,2015A&A...582A..93S} and \citet{2017A&A...607A.126Y}, is essentially the lifetime of molecular clouds and is also the time-scale in which the inter-stellar-medium of a galaxy churns out an optimal or full population of freshly formed embedded clusters, each of which dissolves into the galactic field through gas expulsion, stellar evolution mass loss and two-body relaxation-driven evaporation. The 10~Myr timescale is also consistent with the typical observationally deduced galaxy-wide interstellar medium (ISM) timescale of transforming the ISM via molecular clouds into a new stellar population (\citealt{2004PASJ...56L..45E,2009ApJ...697.1870E}; review: \citealt{2010ARA&A..48..547F,2015ApJ...806...72M}). The disappearance of large molecular clouds around young star clusters also takes about 10~Myr \citep{1989BAAS...21.1189L}.
\end{itemize}

With these axioms, the galaxy-wide stellar initial mass function (gwIMF) can be calculated by adding up the IMF of each embedded star cluster formed in a galaxy in a 10~Myr star formation epoch, according to the ``IGIMF equation'':
\begin{equation}\label{eq:IGIMF}
\xi(t) = \mathrm{d} N_{\star}/\mathrm{d} m=\int_{5~M_\odot}^{10^9~M_\odot} \xi_{\mathrm{\star}}(m,M_{\rm ecl},[Z/X]) \, \xi_{\mathrm{ecl}}(M_{\rm ecl},\bar{\psi}_{\delta t})\,\mathrm{d}M_{\rm ecl},
\end{equation}
where $dN_{\star}$ is the number of stars in the stellar mass interval $m$ to $m+dm$. The integration limits over the embedded cluster mass, $M_{\rm ecl}=5$ and $10^9~M_\odot$ approximately corresponds to the mass of the least and most massive known star-cluster-type system, that is, the smallest stellar groups observed \citep{2012ApJ...745..131K} and the most massive ultra-compact dwarf (UCD, \citealt{2008MNRAS.386..864D}), respectively. Finally, we need to know how the stellar IMF, $\xi_{\mathrm{\star}}$, and the ECMF, $\xi_{\mathrm{ecl}}$, change with physical conditions. These include not only how the slope of the mass functions changes but also how the upper mass limit of the mass function changes according to the calculation given by the optimal sampling theory (see \citealt{2017A&A...607A.126Y} and \citealt[their section 2.1]{2021A&A...655A..19Y}). The adopted $\xi_{\mathrm{\star}}$ and $\xi_{\mathrm{ecl}}$ formulations are achieved and calibrated based on observation. As a side note, it is important to differentiate the random sampling with a provided variable upper stellar mass limit from the optimal sampling resulting in the variable upper stellar mass limit. The relation between the stellar mass upper limit and the embedded cluster mass and the relation between the galactic SFR and the most massive just formed star cluster mass in that galaxy is never plugged into the IGIMF formulation as a known input. They are both observations applied to test the natural outcome of the optimal sampling theory and support the theory.

Since the ``IGIMF equation'' is a function of galaxy-wide gas-phase metallicity, $[Z/X]$, and the average galactic SFR over the $\delta t =10^7$~yr, $\bar{\psi}_{\delta t}$, in units of $M_\odot/$yr, the gwIMF, $\xi$, varies through time according to the galactic metal evolution and SFR fluctuation. The complete set of empirically constrained mathematical expression of how the stellar IMF and the ECMF change with physical conditions is given in \citet[their section 2.1]{2021A&A...655A..19Y} and also in the full version of this thesis. An illustration of the gwIMF variation calculated by the IGIMF theory is demonstrated in \citet[their figure 2]{2018A&A...620A..39J}.

The IGIMF theory is consistent with the observational constraints from the resolved star clusters in and around the MW by construction and is also consistent, automatically with no new parameters introduced, with the observed gwIMF variation of extragalactic galaxies as we will demonstrate below. It has solved a number of previously outstanding extragalactic problems such as explaining the UV extended galactic disks \citep{2008Natur.455..641P} and naturally accounting for the time-scale problem for building up a sufficient stellar population in dwarf galaxies given their low SFRs \citep{2009ApJ...706..516P}.
The theory made the prediction that dwarf galaxies must show a deficit of H$\alpha$ emission relative to UV emission; see \citet{2007ApJ...671.1550P,2009MNRAS.395..394P}. This was verified to be the case by \citet{2009ApJ...706..599L}. The adoption of the IGIMF theory to the galaxy chemical evolution model also leads to good reproduction of what we observe which has previously been unachievable \citep{2015MNRAS.446.3820G,2017MNRAS.464.3812F,2017MNRAS.470..401R,2020MNRAS.494.2355P,2021NatAs.tmp..200M}. Notably, the study of \citet{2020A&A...637A..68Y} on a dwarf galaxy and \citet{2021A&A...655A..19Y} on ETGs of different masses is introduced in Section~\ref{Chapter dwarf} and \ref{Chapter elliptical} below.

In summary, the IGIMF theory, based on an empirical relation of the small-scale IMF variation, is able to account for the IMF of galaxies and explain their special features self-consistently thereby also allowing prediction and is capable of incorporating different empirical constraints on the star-cluster scale if they are well constrained by observations. With this semi-empirical approach, we can constrain the star-cluster-scale IMF and gwIMF and potentially understand the variations of the IMF in different stellar systems.

\subsection{Comparing the IGIMF theory with observation}\label{sec IGIMF-compare}

\begin{figure}
    \center
    \includegraphics[width=13cm]{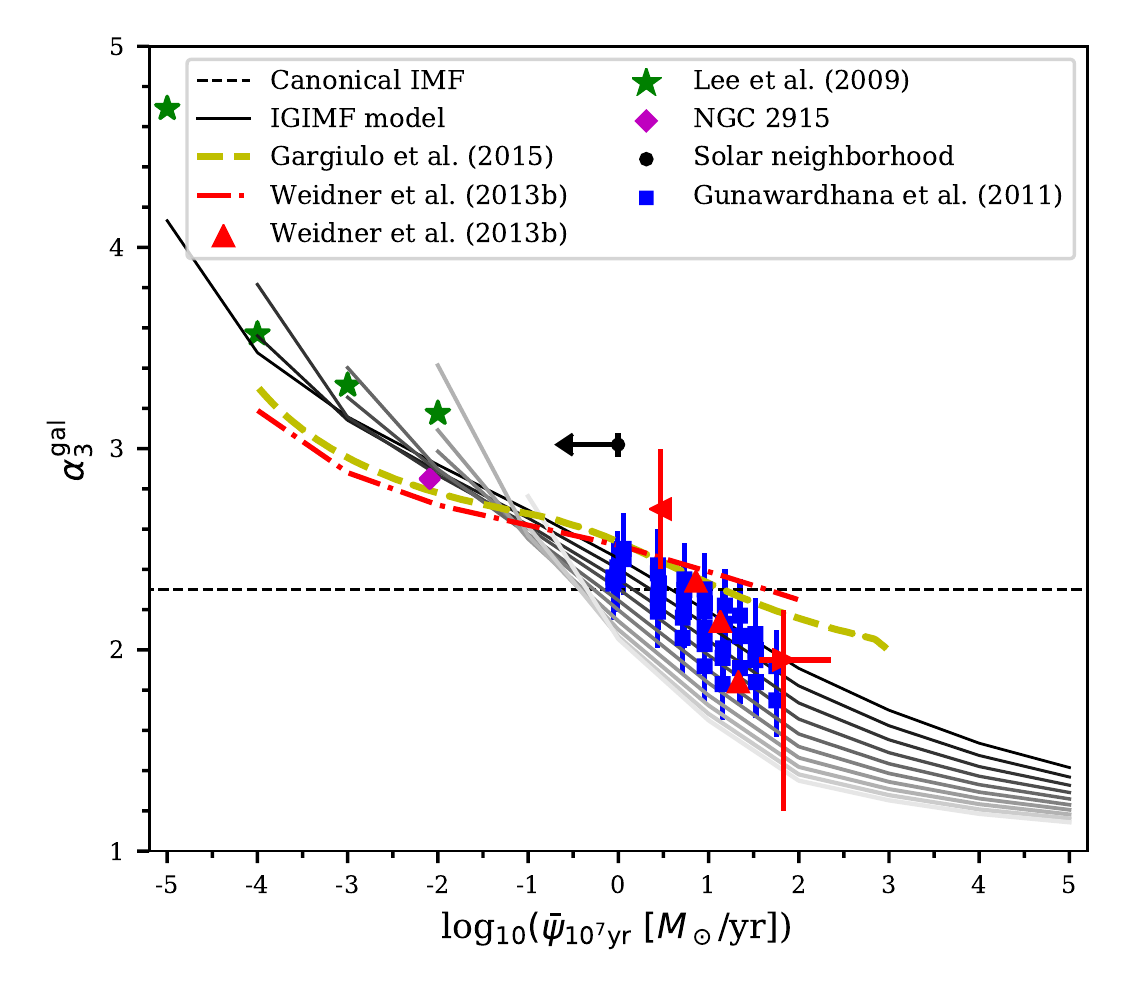}
    \caption{Figure taken from \citealt{2017A&A...607A.126Y}. High mass end power-law index of the galaxy-wide stellar initial mass function (gwIMF), $\alpha_3^{\mathrm{gal}}$ (i.e. the IMF power-law index for stars with mass $m>1$ $M_{\odot}$) as a function of the galaxy-wide SFR. According to the IGIMF theory, $\alpha_3^{\mathrm{gal}}$ values diverge for different SFRs and also vary for different stellar masses. Because at each stellar mass value there exists a different $\alpha_3^{\mathrm{gal}}$--$\bar{\psi}_{10^7\rm yr}$ relation, we plot solid lines for $\log_{10}(m/M_{\odot})=0.2$, 0.4, ..., 2, that is, 1.58, 2.51, ..., 100 $M_{\odot}$ from black to gray (top to bottom) for the IGIMF calculated results. The horizontal dashed line represents the canonical IMF slope for massive stars, that is, $\alpha_3=2.3$. Our theoretical calculation agrees well with observations shown by coloured symbols. See \citet{2017A&A...607A.126Y} for more detail.}
    \label{fig:alpha3SFR}
\end{figure}

The IGIMF theory asserts that the time and space integrated IMF of a galaxy can be calculated from the IMFs in all star clusters ever formed in that galaxy. That is, there are three levels of complexity to compare the theoretical and observational IMF:

\begin{enumerate}
  \item The applied IMF variation law in single star clusters needs to be consistent with the observed stellar populations in different environments. Stellar (luminosity) evolution, binary fraction, and the dynamical evolution of the star cluster need to be accounted carefully \citep{2006MNRAS.373..295P,2013MNRAS.434...84W,2017ApJ...834...94S,2017A&A...607A.126Y,2018MNRAS.481..153O}.
  \item Within a single star formation epoch of about 10 Myr (see the third axiom in Section \ref{sec IGIMF-IGIMF-overview}), the applied embedded cluster mass distribution and its possible variation needs to be consistent with the observation \citep{2004MNRAS.350.1503W,2013ApJ...775L..38R,2017A&A...607A.126Y}. Using the IGIMF theory (Eq.~\ref{eq:IGIMF}) the gwIMF for this single 10 Myr star formation epoch can be calculated and compared with the observed star-forming galaxies where the integrated galactic light is dominated by the newly formed massive stars.
The gwIMF, which is an integration of power-law IMFs, is no longer a power law. Therefore, the slope and the ``power-law index'' of the gwIMF is different for each stellar mass range. We compare the IGIMF calculation with observationally estimated power-law indices of the gwIMF for the massive stars over different stellar mass ranges and galaxy-wide SFRs. For this purpose, the short IGIMF segments in each mass range are approximated by a power-law with index $\alpha_3^{\mathrm{gal}}$. The resulting $\alpha_3^{\mathrm{gal}}$--$\bar{\psi}_{10^7\rm yr}$ relation is shown in Fig.~\ref{fig:alpha3SFR}.
The calculated gwIMF given by the IGIMF theory becomes top-heavy in high-SFR galaxies and top-light in low-SFR galaxies, matches the observations nicely.
The significance of this test is that the IGIMF formulation is able to explain the galaxy-wide IMF variation purely from the observationally determined star-cluster scale IMF variation (i.e. from \citealt{2012MNRAS.422.2246M}), showing that the IMF behaviour on large and small scales are linked by a unified theory. This is a consistency success of the IGIMF theory (regarding the massive stars).
  \item The present-day galactic IMF is a composition of multiple stellar populations formed at different times, that is, a time-integrated gwIMF (TIgwIMF). Since the gwIMFs are different for each 10 Myr star formation epoch depending on the instantaneous galactic environment, we need to apply a self-consistent galaxy evolution model (chemical or even chemo-hydrodynamical model) to calculate the TIgwIMF. This can then be compared with galaxies that have stopped their star formation for a certain time which allows observational tests for lower-mass stars. The TIgwIMF also determines the mass-to-light ratio of a galaxy which can be constrained by the dynamical modelling or lensing studies of galaxies. To this end, a self-consistent galaxy chemical evolution model is developed as is introduced in the next section.
\end{enumerate}

~~

\begin{tcolorbox}
The IGIMF theory is a mathematical framework that links the IMF on different spatial scales. By improving the description of the small embedded star cluster scale IMF variation, the large galactic scale IMF variation calculated according to the IGIMF theory will become more accurate.
\end{tcolorbox}

~~

%% file: 03.tex
\section{Galaxy chemical evolution model}\label{sec GCE}

\subsection{Introduction}

The observed stellar abundance ratios in galaxies with different masses, SFRs, and metallicities has not been well explained by previous galaxy chemical evolution (GCE) models assuming the invariant canonical IMF.
GCE studies of massive elliptical galaxies have suggested a problem in simultaneously reproducing the total metallicity and $\alpha$-enhancement of them \citep{2017MNRAS.466L..88D,2017MNRAS.464.4866O} indicate the necessity of introducing at least one new degree of freedom, for example, a varying and top-heavy gwIMF \citep{2010MNRAS.402..173A,2018MNRAS.475.3700M}.
Given that the evolution of galaxies sensitively depends on the IMF and a growing evidence that suggest a systematic variation of the IMF as is successfully described by the IGIMF theory, the IGIMF theory is likely to be the missing link of the problem and needs to be explored. 

When the gwIMF varies, GCE will differ from the canonical estimate. However, most GCE models were developed initially assuming an invariant IMF and were then modified to adopt a variable IMF. It is not always the case that this paradigm shift is done or explained correctly. In addition, the previous works introduce ad-hoc variations of the IMF while it is crucial to apply the IMF variation formulation that is consistent with established observational constraints to result in a meaningful conclusion.
Not only that most of the GCE code are not publicly available and the few open-source GCE code (e.g. NuPyCEE, \citealt{2016ascl.soft10015R}, Chempy, \citealt{2017A&A...605A..59R}, flexCE, \citealt{2017ApJ...835..224A}, and VICE \citealt{2020MNRAS.498.1364J}) all assume invariant IMFs but, more importantly, the time structure of the code needs to be redesigned to incorporate the observationally tested IGIMF theory (see Section~\ref{sec GCE-Structure} below), the difference in the time structure is detailed in the full version of this thesis) such that it is not as straightforward as one might think to \textit{correctly} implement the IGIMF theory on existing GCE codes.

For this contribution, we develop a new open-source Python3 code, GalIMF, that is able to couple an environment-dependent IMF theory with the GCE. The instantaneous environment (e.g. gas metallicity) determines and constantly updates the gwIMF while the gwIMF of each stellar population ever formed determines the later metal enrichment process, resulting in a non-linear IGIMF effect. GalIMF stands for the Galaxy-wide Initial Mass Function. GalIMF version 1.0 (with a companion paper \citealt{2017A&A...607A.126Y}) is a Python 3 module that computes gwIMFs based on the IGIMF theory. Later, GalIMF version 1.1 (with a companion paper \citealt{2019A&A...629A..93Y}) couples the IGIMF theory with galaxy chemical evolution.

The development of a fully functional GCE code is a technical and demanding task. In this summary, we only outline the basic structure of the model. For more details and the performance of the code, please check \citet{2019A&A...629A..93Y}, the full version of this doctoral thesis, and our online repository in the grey box below.

~~

\begin{tcolorbox}
The GalIMF code, deployment manual, examples, as well as version update records, are available at GitHub:
https://github.com/Azeret/galIMF
\end{tcolorbox}

~~

\subsection{Structure of the GalIMF code}\label{sec GCE-Structure}

\begin{figure}
        \centering
        \includegraphics[width=11cm]{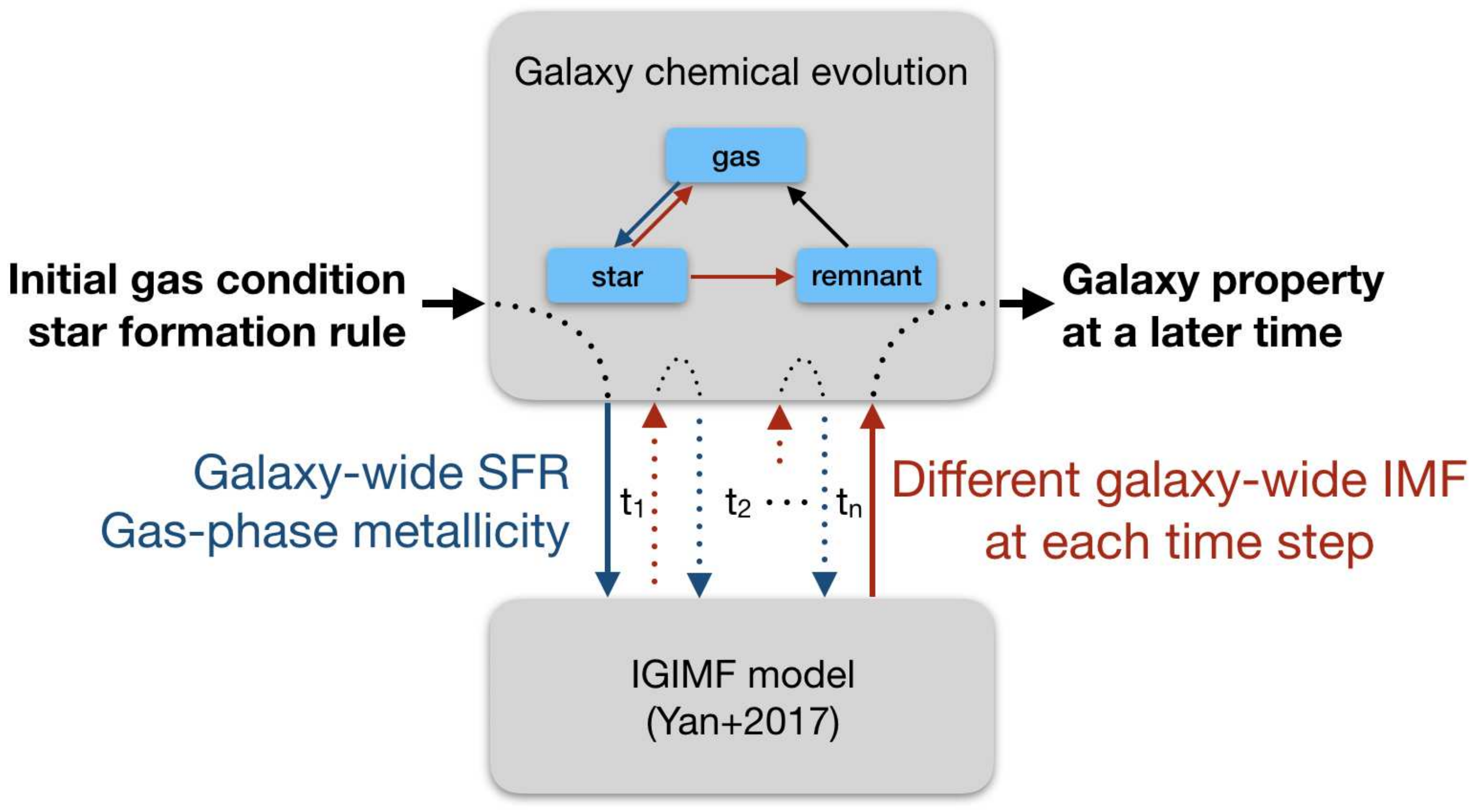}
        \caption{A schematic diagram for the GalIMF code structure, indicating the interaction between the GCE module and the IGIMF calculation module and the input/output of the code. The gwIMF is calculated according to the IGIMF theory for each 10~Myr star formation epoch. See \citep{2019A&A...629A..93Y} for more details.}
        \label{fig:structure}
\end{figure}

The structure of our GCE code is shown in Fig.~\ref{fig:structure}. There are three basic mass component to be considered in the model: gas, stars, and stellar remnants. The abundance evolution of different elements in the gas and stars are recorded separately. Population of stars form according to the assumed star formation law (e.g. the Kennicutt-Schmidt law) and the IGIMF theory and possess the same abundance as the abundance of the gas that it forms from. 
Other than applying different star formation laws, it is also possible to specify the star formation activity for each 10 Myr time step. That is, fix the SFH of a galaxy before the simulation starts. This mode has the advantage to compare different IMF formulations for identical SFHs (e.g. \citealt{2019A&A...629A..93Y}).
To incorporate the environment-dependent gwIMF into the GCE model, the GCE module requires the gwIMF calculation module to calculate the gwIMF according to the IGIMF theory at each time step according to the instantaneous galaxy-wide SFR and gas-phase metallicity. Different from the invariant IMF assumption where the star formation activity is assumed to be a continuum (i.e. the model can form a fraction of a star and explode a fraction of a supernova if given a short time step, having the benefit to easily integrate the stellar yield), the calculation of gwIMF with the IGIMF theory is done for 10 Myr time steps asserting that these clusters belong to the same correlated star formation epoch and populate an optimally-sampled ECMF (see Section~\ref{sec IGIMF-IGIMF-overview}).

When the stellar lifetime, which is a function of stellar mass and metallicity, is exhausted, the star becomes a stellar remnant with a large portion of its mass ejected into the gas phase through stellar wind or supernova explosion. Precise stellar ages coupled with a short evolution time step (10 Myr) is implemented to result in accurate abundance evolution histories. The element yield of a dying star depends on its initial mass and metallicity and several different stellar yield table has been implemented (e.g. \citealt{1995ApJS..101..181W}, \citealt{1998A&A...334..505P}, \citealt{2001A&A...370..194M}, and \citealt{2006ApJ...653.1145K}). The stellar remnants can still participate in the galaxy chemical evolution by triggering a type Ia supernova (SNIa). The SNIa rate depends on the number of potential progenitors (i.e. stellar remnants within certain mass range) and the delay time distribution of the SNIa (the power-law DTD in \citealt{2012PASA...29..447M}).
Since a variable gwIMF is applied, we re-calibrated the number of SNIa to account for the IMF variation and potential environmental effects (detailed below).
The SNIa yield is adopted from \citet[their TNH93 dataset]{1997MNRAS.290..623G} and does not depend on the property of the SNIa progenitor. Thus, the gas phase is gradually enriched from a primordial element composition, adopted from \cite{2016RvMP...88a5004C}, to higher and higher metallicity. 
At each time step when the gas element abundance is to be updated, the yields from stars and SNIa, that contributes between the current and the previous time step, for all previously formed stellar populations are integrated (therefore, the computational time of the code is proportional to the square of the number of the 10~Myr star formation epochs). 
All elements in the gas reservoir are assumed to be always well mixed, that is, there is only a single gas phase and this is called a single-zone GCE model. More technical details are provided in \citet{2019A&A...629A..93Y}. Galactic gas inflow and outflow can be modelled by our code which adds in or remove gas from the gas reservoir. 

Finally, the code is able to output the mass evolution of different mass component and elements, the rate of supernovae, and the gwIMF evolution at all time steps. The average element abundance of the galactic stellar population at a given time is given both by a mass-weighted and a stellar luminosity-weighted value.

\subsection{The SNIa production efficiency}\label{Chapter GCE sec SNIa}

~~

\begin{tcolorbox}
The number of SNIa depends on the number of possible SNIa progenitors, that is, on the number of stars with a mass between about 3 to $8 M_\odot$, therefore, on the gwIMF.
\end{tcolorbox}

~~

\begin{figure}
    \centering
    \includegraphics[width=10cm]{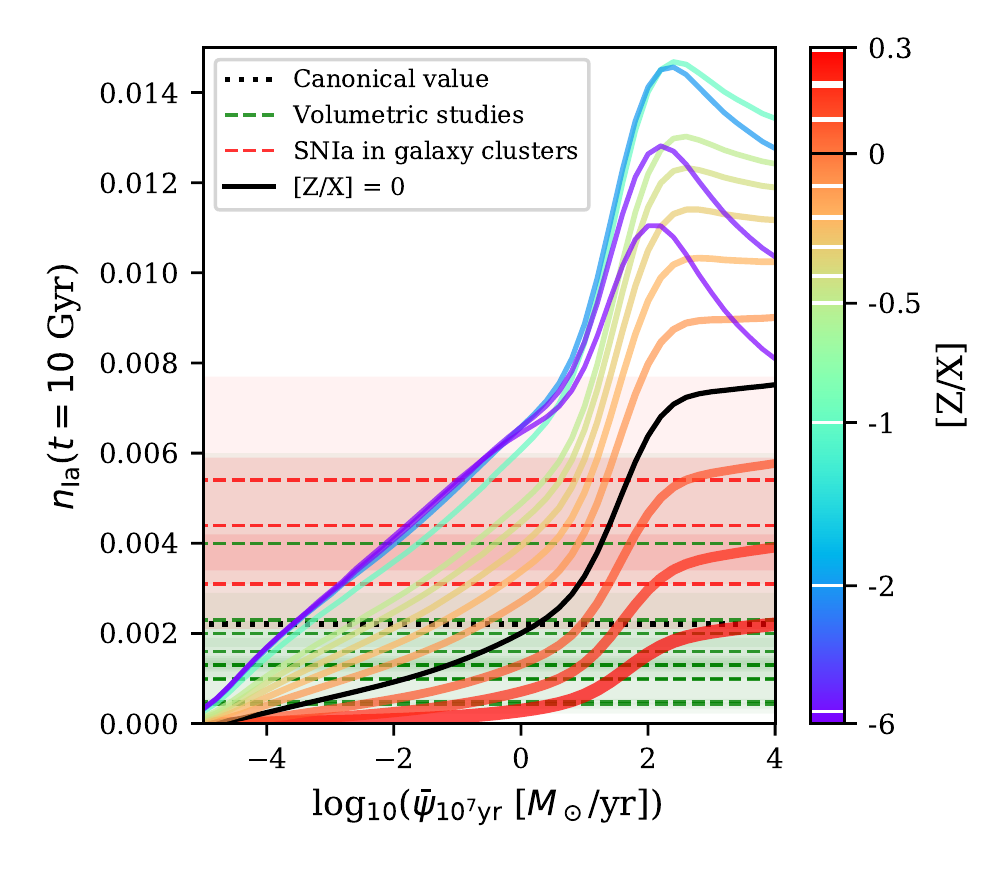}  
    \caption{Figure taken from \citealt{2021A&A...655A..19Y}. The number of SNIa per unit stellar mass formed, $n_{\rm Ia}$, after $t=10$~Gyr since the formation of the stellar population, for an elevated SNIa formation efficiency for high-SFR galaxies as is described by Eq.~\ref{eq:error function}. The $n_{\rm Ia}$ of a galaxy depends on the galaxy-wide IMF which is given by the IGIMF theory (Eq.~\ref{eq:IGIMF}) as a function of the galaxy-wide SFR, $\bar{\psi}_{10^7\rm yr}$, and metallicity, $[Z/X]$. The black line is the relation for $[Z/X]=0$. Other lines with different colours represent different values of $[Z/X]$ as indicated by the white stripes on the colour map on the right: $[Z/X]=0.3$, 0.2, 0.1, -0.1, -0.2, -0.3, -0.4, -0.5, -1, -2, -4, and -6. The black horizontal dotted line represents the canonical $n_{\rm Ia}(t=10~\mathrm{Gyr}, \xi_{\mathrm{canonical}})$ value of $0.0022/M_\odot$ \citep{2012PASA...29..447M}. The green and red horizontal dashed lines indicate observational constraints on $n_{\rm Ia}(t=10~\mathrm{Gyr})$ for SNIa surveys up to a certain redshift and in galaxy clusters, respectively, suggesting that clustered galaxies have a higher $n_{\rm Ia}$ than field galaxies. The shaded regions represent the uncertainty ranges of the horizontal dashed lines.}
    \label{fig:SNIa_variation_IGIMF_varKappa}
\end{figure}

The total number of SNIa explosions for a simple stellar population (SSP, stars formed at the same time with the same metallicity) after $t$ years of the birth of the SSP per unit stellar mass of the SSP (i.e. the time-integrated number of SNIa per stellar mass formed until time $t$) is
\begin{equation}\label{eq:SNIa rate}
    n_{\rm Ia}(t, \xi, \bar{\psi}_{\delta t})=N_{\rm Ia}(\xi, \bar{\psi}_{\delta t}) \int_{0}^tf_{\rm delay}(t)\mathrm{d}t \; ,
\end{equation}
where $f_{\rm delay}$ is the delay time distribution (DTD) function, that is, the fraction of exploded SNIa for an SSP with age $t$, and $N_{\rm Ia}$ is the SNIa production efficiency, that is, the total number of SNIa (for $t=\infty$) per unit mass of stars formed in the SSP.

The SNIa production efficiency, $N_{\rm Ia}$, is calculated as
\begin{equation}\label{eq:N_Ia}
    N_{\rm Ia}(\xi, \bar{\psi}_{\delta t}) = \frac{n_{3,8}(\xi)}{M_{0.08,150}(\xi)} \cdot B_{\rm bin}(\bar{\psi}_{\delta t}) \cdot \frac{n_{3,8}(\xi)}{n_{0.08,150}(\xi)} \cdot C_{\rm Ia}(\bar{\psi}_{\delta t}),
\end{equation}
where $n_{3,8}$ and $n_{0.08,150}$ are the number of stars within the mass range given by their subscript (in the unit of $M_\odot$). Similarly, $M_{0.08,150}$ is the mass of stars in the mass range indicated by its subscript. Therefore the first and third terms on the r.h.s. depend on the IMF. $B_{\rm bin}$ denotes the binary fraction of stars within the mass range of 3 to $8~M_\odot$ and $C_{\rm Ia}$ is the realisation probability of an SNIa explosion for the potential SNIa progenitor system. Both depend on environmental factors such as the stellar density and metallicity. The variation of the SNIa production efficiency has been suggested by, for example, \citet{2018MNRAS.479.3563F} and \citet{2021MNRAS.502.5882F} who find a significantly enhanced occurrence of SNIa in galaxy clusters. A more detailed interpretation and comprehensive discussion of the above equation is given in \citet{2021A&A...655A..19Y}, where this has been published for the first time.

In addition to the IMF variation effect, we define an SNIa realisation re-normalisation parameter, $\kappa_{\rm Ia}(\bar{\psi}_{\delta t})$, to account for the environmental variation on the SNIa production efficiency.
$\kappa_{\rm Ia}$ represents the variation of the overall SNIa realisation parameter as a function of the galaxy-wide SFR,
\begin{equation}\label{eq:SNIa_renormalization}
    \kappa_{\rm Ia}(\bar{\psi}_{\delta t}) = \frac{B_{\rm bin}(\bar{\psi}_{\delta t}) \cdot C_{\rm Ia}(\bar{\psi}_{\delta t})}{B_{\rm bin}(\psi_0) \cdot C_{\rm Ia}(\psi_0)},
\end{equation}
which allows the variable $B_{\rm bin}(\bar{\psi}_{\delta t}) \cdot C_{\rm Ia}(\bar{\psi}_{\delta t})$ to become larger in massive galaxies.

Through trial and error (see Section~\ref{Chapter elliptical} and \citealt{2021A&A...655A..19Y}), we find that assuming
\begin{equation}\label{eq:error function}
    \kappa_{\rm Ia}(\bar{\psi}_{10^7\rm yr}) = 1.75 + 0.75 \cdot \mathrm{erf}[\mathrm{log}_{10}(\bar{\psi}_{10^7\rm yr}~[M_\odot/\mathrm{yr}]) \cdot 1.25 - 2],
\end{equation}
where erf stands for the Gauss error function, which increases the SNIa production efficiency for high-SFR galaxies up to a factor of $1.75+0.75=2.5$, leads to a result that roughly fits the observed $\tau_{\rm SF, SPS}$--$M_{\rm dyn}$ relation suggested by \citet{2015MNRAS.448.3484M}. The required higher SNIa production efficiency for massive ETGs (given by our GCE model, \citealt{2021A&A...655A..19Y}, see Section~\ref{Chapter elliptical}) agrees beautifully with the independent theoretical expectation \citep{2002ApJ...571..830S} and observation \citep{2021MNRAS.502.5882F} for the first time.

Since the gwIMF predicted by the IGIMF theory is a function of the galaxy-wide SFR and metallicity (see Eq.~\ref{eq:IGIMF}), the number of SNIa per unit stellar mass integrated over 10 Gyr formed, $n_{\rm Ia}(t=10~\mathrm{Gyr})$, changes as a function of these two parameters as is shown by Fig.~\ref{fig:SNIa_variation_IGIMF_varKappa}.

\subsection{Galaxy evolution results}\label{sec GCE-Galaxy evolution results}

\begin{figure}
    \centering
    \includegraphics[width=10cm]{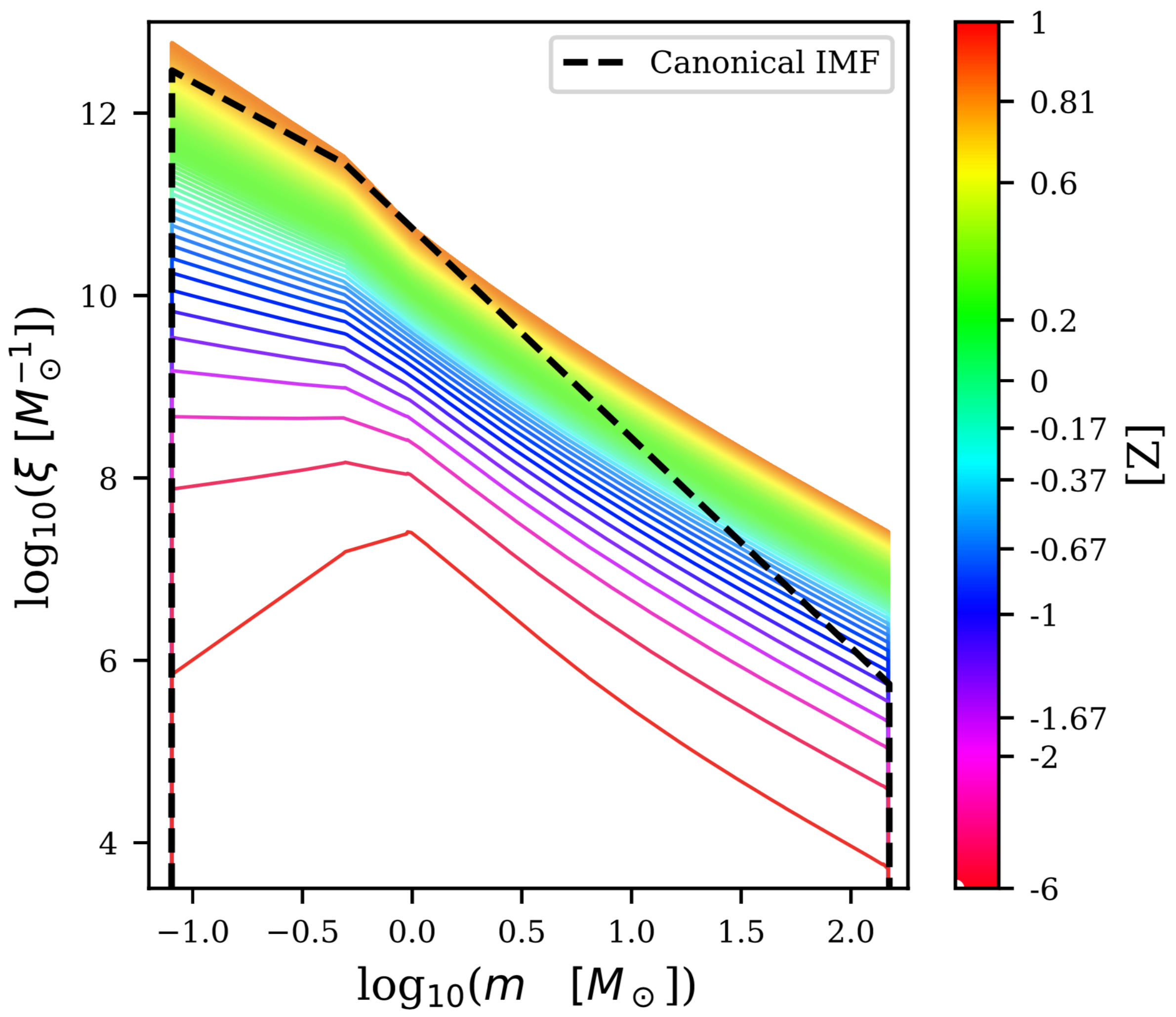}
    \caption{Figure taken from \citealt{2019A&A...629A..93Y}. Evolution of the time-integrated galaxy-wide initial mass function (TIgwIMF) for a galaxy with a log-normal SFH, where $\xi$ is the total number of stars in the galaxy within a unit mass range. The solid lines represent the TIgwIMF after 10, 20, ..., and 1000 Myr since the beginning of the star formation (from bottom to top). The colour of the lines indicates the metallicity $[Z]$ in gas at the time (therefore also the initial metallicity of the stars formed at that time). The dashed line is the canonical Kroupa IMF, normalized to have the same $\xi$ as the final TIgwIMF at $m=1~M_\odot$.}
    \label{fig:1000_IMF_evolution}
\end{figure}

The GalIMF GCE code has been tested by reproducing previous publication results assuming the canonical IMF \citep{2019A&A...632A.110Y,2020A&A...637A..68Y}. The evolution of mass, number of supernovae, and element abundance are demonstrated in the full version of this doctoral thesis and in, for example, \citet{2019A&A...629A..93Y}. When adopting the IGIMF theory, the gwIMF of each 10~Myr time step is metallicity- and SFR-dependent, therefore, the time-integrated gwIMF (TIgwIMF) also evolves with time.
Figure~\ref{fig:1000_IMF_evolution} demonstrates the TIgwIMF for a log-normal SFH calculated by our GalIMF code applying the IGIMF theory, changing gradually from top-heavy bottom-light (bottom lines) to top-heavy bottom-heavy (top lines).

The gwIMF is top-heavy at early times of the evolution due to the high SFR during the early starburst (this being directly related to the rapid formation of supermassive black holes, \citealt{2020MNRAS.498.5652K}) and the gwIMF for the low-mass star evolves from bottom-light to bottom-heavy due to metal enrichment \citep[their eq. 4]{2021A&A...655A..19Y}. See a more in-depth discussion in \citealt{2018A&A...620A..39J}).

~~

\begin{tcolorbox}
For the old ETGs, the observations of the still living low-mass stars do not give information on the gwIMF slope for the massive stars. We demonstrated here that TIgwIMF can be both top-heavy and bottom-heavy. It would be incorrect to assume a single power-law IMF for such galaxies and to suggest a top-light gwIMF due to an observed bottom-heavy gwIMF.
\end{tcolorbox}

~~

%% file: 04.tex
\section{Applications of the GalIMF model}\label{Chapter application}

\subsection{Chemical evolution of dwarf galaxies}\label{Chapter dwarf}

The nearby dwarf galaxies provide a unique opportunity to study their gwIMFs. Firstly, many of these low mass galaxies are formed with a short starburst, then the star formation stops due to a depletion of gas. This leaves a relatively simple stellar population with no recently formed younger stars. Not only the SPS is more reliable, but the SFH in the GCE model can also be greatly simplified. Secondly, there are a large number of them relative to massive galaxies, providing more close-by targets within which we can measure the abundances of single stars with high accuracy. This measurement is important as the abundances of stars formed at different times trace the gas abundance evolution of the galaxy. \citet{2018ApJ...852...50F} has suggested that the GCE model is able to deduce whether a difference in the abundance evolution is caused by a SFH or IMF variation. Indeed, the GCE of dwarf galaxies is different from the MW's. For instance, \citet{2020A&A...642A.176T} find that the stellar $[\alpha/Fe]$ values of Sextans, Sculptor, and Fornax dwarf galaxies bend and decrease at different $[Fe/H]$ values than in the MW. \citet{2021ApJ...910..114M} find, in addition, that different $\alpha$-elements behave differently as one would expect if the gwIMF varies because different elements are produced mainly by stars of different mass.

In \citet{2020A&A...637A..68Y}, we apply the IGIMF theory to the GCE modelling of one of the best-observed ultra-faint dwarf (UFD) satellite galaxies, Bo\"otes~I. We find that the $[\alpha/Fe]$--$[Fe/H]$ relation of Bo\"otes~I can be well reproduced if it has the average gas-depletion timescale of dwarf galaxies \citep{2009ApJ...706..516P}, challenging the idea that UFDs have a much longer gas-depletion timescale than dwarf galaxies. Remembering that when the galaxy stellar mass as an observable is known, the gas-depletion timescale determines the average SFR which then determines the gwIMF of the massive stars that affect the shape of the $[\alpha/Fe]$--$[Fe/H]$ relation. Therefore, the IGIMF theory suggesting a top-light IMF naturally accounts for the observed $[\alpha/Fe]$--$[Fe/H]$ relation of Bo\"otes~I given its low mass and SFR. The recent GCE study on LMC and NGC 2005 agrees with our result \citep{2021NatAs.tmp..200M}.

The parameters applied in our GCE model are not fine-tuned to force a good fit. They are either inherited from previous studies or significantly constrained by independent sources or data. It is remarkable that the UFD Bo\"otes~I is so well described using the IGIMF theory, better than the best-fit model assuming the canonical IMF (compared in \citealt[their table 1]{2020A&A...637A..68Y}). This confirms that the optimal sampling of ECMF and the IGIMF theory is most likely correct.

~~

\begin{tcolorbox}
The gas-depletion timescales of galaxies are estimated by GCE studies and the conclusion depends on the assumed IMF. While the studies assuming the canonical IMF suggest a long gas-depletion timescale of the UFDs, our study assuming the IGIMF theory leads to a normal gas-depletion timescale which is, in fact, the same as the average gas-depletion timescale of dwarf galaxies (cf. \citealt{2009ApJ...706..516P}), suggesting that UFDs are an extension of the ordinary dwarf galaxies to lower stellar masses instead of a separated species (cf. \citealt{2021ApJ...920...92J}).
When exploring a new IMF law, one must keep in mind that galactic properties such as the star formation law determined previously from studies assuming the canonical IMF are no longer correct.
\end{tcolorbox}

~~

In addition, we point out in \citet{2020A&A...637A..68Y} that the GCE model can be used to constrain the IMF variation law of low-mass stars and apply this method for the first time. Once the SFH and the gwIMF of the massive stars are determined by the $[\alpha/Fe]$--$[Fe/H]$ relation, which is not possible previously without the abundance measurements of individual stars in the dwarf galaxies, and in addition, by an independent SPS study \citep{2014ApJ...796...91B} of the galaxy, the mean stellar metallicity of the galaxy is affected by the gwIMF of the low-mass stars. Since the mean stellar metallicity is also deducible from SPS, the variation of the gwIMF of the low-mass stars can be constrained. Our estimated low-mass IMF variation law \citep[their eq. 9]{2020A&A...637A..68Y} is consistent with independent estimation based on Galactic GCs \citep{2002Sci...295...82K} and our methodology is independently reproduced by \citet{2021MNRAS.503.6026R} resulting in the same conclusion.

For more details, we refer the reader to \citet{2020A&A...637A..68Y}.

\subsection{Evolution of ETGs with different masses}\label{Chapter elliptical}

GCE is a useful constraint on the SFH of galaxies. The ratio of CCSN and SNIa explosions decreases with time after the formation of a stellar population, therefore, also the $\alpha$ elements-to-iron peak elements ratio (e.g. [Mg/Fe]) of the galactic gas. A shorter or longer star formation timescale of a galaxy then results in a higher or lower stellar average [Mg/Fe] value, respectively. In this section, we run the GCE calculation for ETGs and study how the newly implemented IGIMF theory affects our understanding of the formation history of ETGs. 

The SFH of a galaxy is not easy to determine when we can only look at the present-day snapshot of it, in particular for the galaxies that cannot be resolved and examined with a CMD of its stars. 
In this case, the standard way to estimate the SFH, given the integrated galactic light, is to fit an artificially synthesized stellar population with a \textit{realistic} (see below) distribution of age, metallicity, and mass to the observed spectrum. This is the so-called stellar population synthesis (SPS) method. One problem of the SPS method is that the solution is not unique. The distribution of all three properties of the stellar population can be adjusted to fit the data. For example, a more fluctuating SFH instead of a smooth SFH \citep{2021arXiv210904488F}, a stellar population with some elements more abundant than usual, and a different IMF such as the one given by the IGIMF theory. All these factors affect the final estimation. 
One of the important requirements, for the synthesized stellar population to be \textit{realistic}, is that the age and metallicity distribution can be naturally reproduced by the GCE model. That is, the stellar population becomes more enriched as time passes due to the previous populations.

A combination of SPS and GCE models to fulfil this requirement has been demonstrated recently by \citet{2020MNRAS.498.5581B}. Historically, the SPS method does not consider GCE and only estimates the average age and metal abundance of the stars. The formation time scale of the stars, SFT, is estimated on top of that using the $\alpha$ elements-to-iron peak elements ratio. This ratio is sensitive to the SFT because most of the two groups of elements are produced by different types of supernovae. $\alpha$-elements are almost exclusively produced by CCSN over a short timescale, while about half of the iron-peak elements are continuously produced by thermonuclear supernovae long after the formation of the stellar population. With this method, the most famous ``downsizing'' relation of ETG formation is established by \citet{2005ApJ...621..673T} and \citet{2010MNRAS.404.1775T} with the result that more massive ETGs form at an earlier time with a shorter SFT. This conclusion revolutionized our understanding of how these galaxies form at the time when the standard cosmology model (LCDM) expected that more massive galaxies formed at the centre of massive dark matter halos (DMHs) as a result of the hierarchical merging of smaller DMHs and therefore should have formed later over longer timescales.

With the intention to introduce the IGIMF theory into the previous calculation done by Thomas, we first reproduced the previous ``downsizing'' relation in \citet{2019A&A...632A.110Y} with the GalIMF code. We also improved the SFT estimation of the GCE method by considering the constraints of mean stellar metallicity in addition to the $\alpha$ elements-to-iron peak elements ratio. This is because the $\alpha$-to-iron ratio yields are different for stars with different metallicity. Massive stars with a higher initial metallicity have a lower [Mg/Fe] yield because the stronger stellar wind of these stars mainly reduces the $\alpha$ element production \citep{2012ceg..book.....M}.

We point out a few difficulties of the canonical ``downsizing'' scenario. Namely, the idea of shorting the SFT to reproduce the observed $\alpha$-element enhancement only works in a simple GCE model where stellar ejections mix with environmental gas instantaneously, but does not work in a more realistic hydrodynamical model of galaxies. When considering the cooling and mixing of the gas, a non-negligible time is required for the stellar ejection to participate in the formation of a new generation of stars and increase the mean stellar metallicity. It happens that when the SFT is too short, the mean stellar metallicity of the galaxy does not have time to enrich to the observed value. Therefore, people struggle to fit both the $\alpha$-element enhancement and mean stellar metallicity simultaneously as is demonstrated and explained in \citet{2017MNRAS.466L..88D} and \citet{2017MNRAS.464.4866O}. The same intrinsic problem persists in \citet{2018MNRAS.479.5448B} and \citet{2019ApJ...880..129P}. In addition to the nearby quiescent galaxies, gravitational lensing observations of high-redshift quiescent galaxies find that the metal abundance and $\alpha$-elements are both enriched and that this cannot be explained by simple GCE models that vary only the SFT \citep{2020ApJ...897L..42J}. In fact, the iron and $\alpha$-element rich galaxy is an analogy but a reversed problem we have seen in the above section discussing the dwarf galaxies that are both iron and $\alpha$-element poor compared to the MW \citep{2021NatAs.tmp..200M}. 

We suggest that the same IGIMF formulation constitutes a natural solution to all the above difficulties.
In the follow-up paper, \citet[hereinafter Yan21]{2021A&A...655A..19Y}, we perform the same SFT calculations but by introducing the environment-dependent IMF to this problem for the first time. We noted that there is no surprise if one manages to fit the same observational constraints with an arbitrarily adjusted IMF. The point of our work is to test the independently developed and observationally tested IGIMF theory to get a reliable insight into how the ETGs form. By doing this, we found out that the $\alpha$-element production in the massive ETGs outperforms the iron-peak elements to a great extend, altering the SFT estimation significantly (Yan21, their Fig. 4). For the result to be consistent with other SFT constraints, such as using SPS methods (Yan21, their Fig. 7), it necessarily follows that the iron-element production is also elevated in these massive ETGs, most likely due to a higher production efficiency of SNIa in the high stellar density region of the high-SFR massive galaxies. We estimated how much the production efficiency of SNIa needs to rise as a function of galaxy mass (Yan21, their Fig. 5) and compared our estimation with the observational constraints given by direct SNIa number counts in field galaxies and in galaxy clusters (Yan21, their Table 1 and Fig. 6). As mentioned in Section~\ref{Chapter GCE sec SNIa}, we find that our model nicely agrees with the theoretical expectation from the N-body simulations and the observational suggestion that the production efficiency of SNIa in higher density regions appears to be higher. This is exactly why the canonical IMF model cannot work as it would require a lower SNIa production efficiency for massive galaxies to better fit with the stellar abundance ratios, being inconsistent with the observations. Finally, we calculated the dynamical mass-to-light ratios (M/L) of the ETGs in our best-fit model and found them to naturally reproduce the observed M/L--galactic-mass relation (Yan21, their Fig. 10). The bottom-heavy gwIMF of massive ETGs in our model (Yan21, their Fig. 9) also agrees with recent observational suggestions discussed in Section~\ref{sec IGIMF-Observation}. 

~~

\begin{tcolorbox}
In summary, many observed ETG properties, such as their element abundance evolution, SNIa production efficiency, and M/L, from dwarf to the most massive ETGs, come together naturally and are straightforwardly explained, all are consequences of the application of the IGIMF theory to the monolithic collapse of post-Big Bang gas clouds.
\end{tcolorbox}

~~

For more details, we refer the reader to \citet{2019A&A...632A.110Y} and \citet{2021A&A...655A..19Y}.

\subsection{Other applicaitons}

The applications of the IGIMF include but not limit to the following works with me as a coauthor:

\begin{itemize}
  \item The formation of star clusters (\citealt{2018A&A...612A..74K} and \citealt{2021MNRAS.506.4131W});
  \item The formation of galaxies and their corrected SFR \citep{2018A&A...620A..39J};
  \item The corrected cosmic star formation history \citep{2020A&A...636A..10C}.
\end{itemize}

Please see the references and attached publication list for more detail.

%% file: 05.tex
\section*{Summary and conclusion}\label{Chapter conclusion}
\addcontentsline{toc}{section}{Summary and conclusion}

A systematic IMF variation is currently the simplest explanation to consistently account for the vast variety of anomalies observed for stellar populations on different scales and at different redshifts (Section~\ref{sec IGIMF-Observation}). With the empirically calibrated IGIMF theory describing how the IMF varies \citep{2017A&A...607A.126Y} and the computational tool developed specifically for this study \citep{2019A&A...629A..93Y}, we are able to explore the effect of a realistic IMF variation scenario on the chemical evolution of different types of galaxies. From the dwarf galaxy with a stellar mass of only a few times $10^4~M_\odot$ to the most massive monster galaxies with a dynamical mass of $10^{12}~M_\odot$, our GCE model adopting the IGIMF theory is able to explain the individual stellar chemical abundances inside the galaxy better than models that assume the IMF is invariant.

In \citet{2020A&A...637A..68Y}, it is shown that the IGIMF theory is able to naturally explain the element ratio of the dwarf galaxy Bo\"otes~I. According to our best-fit solution, we find that Bo\"otes~I is not unusual but follows the same gas-consumption process as other galaxies. This practice also allows us to put constraints on the low-mass IMF dependence on the metallicity for the first time. This is an important result because the IMF variation law of low-mass stars outside our Galaxy is difficult to constrain. The observed relation from old quiescent galaxies, where massive stars do not outshine the low-masses, is for the time-integrated present-day stellar mass function that is different from the varying IMF.

In \citet{2019A&A...632A.110Y} and \citet{2021A&A...655A..19Y}, we explain the scaling relation between the abundance ratios and galaxy mass which has not been understood in canonical models. \citet{2019A&A...632A.110Y} demonstrated the importance of considering stellar metallicity and $\alpha$-element enhancement together when estimating the star formation timescale of a galaxy. The results show that the model adopting the canonical IMF does not give formation timescales that are consistent with independent constraints even if one allows a tuning of the stellar element yield table. Either the formation timescales for massive galaxies are too short or the formation timescales for the low-mass galaxies are too long. Following this work, \citet{2021A&A...655A..19Y} shows that the IGIMF theory provides a natural solution to this problem. The more top-heavy galactic IMF of the massive and high-SFR galaxies produces a lot more $\alpha$-elements at early times, modifying the estimated stellar formation timescales of these massive galaxies to longer values. In fact, the empirically calibrated IGIMF formulation predicts that too many $\alpha$-elements should have been produced such that more iron-peak elements should also have been produced to result in the observed $\alpha$ element-to-iron peak element ratio. Therefore, we are able to use this constraint to estimate how the production efficiency of SNIa should change in a different environment.

We conclude that the realistic IMF variation can have a strong impact on the modelled GCE and our interpretation of galaxy observations. The estimated galaxy mass and formation timescales are different from the canonical model, which affects our understanding of how the universe evolves. The abnormal element abundances, if accounted by the IMF difference, no longer requires an adjustment of the stellar evolution model or the introduction of exotic stellar explosion scenarios. Therefore, the IMF variation, being the missing link between the stars and the universe in many of today's astronomical studies, needs to be taken into consideration.

The future development of the field will require a significant amount of effort to further test and constrain the IMF variation formulation empirically in order to give an accurate and reliable description of how the IMF varies. Usage of different GCE codes with different advantages that make possible cross-checking are highly encouraged. Open-source codes or, at least, an extensively detailed description of the GCE method, should become the standard. Communication and collaborative effort are most valuable.